\documentclass[12pt]{article}
\usepackage{amsmath,amssymb,latexsym}

\begin{document}

\begin{flushright}
	YITP-03-11 \\
	hep-th/0303240
\end{flushright}

\vspace{2.5cm}

\begin{center}
{\Large\bf Unitary-Matrix Integration on 2D Yang-Mills Action}\\
\bigskip
\vspace{1.5cm}
{\large Yoshinobu H}ABARA~\footnote{e-mail: habara@yukawa.kyoto-u.ac.jp}\\
\vspace{0.5cm}
{\it  Yukawa Institute for Theoretical Physics,}\\
{\it  Kyoto University, Kyoto 606-8502, Japan}
\end{center}

\vspace{1cm}

\begin{abstract}
Using the idea of Itzykson-Zuber integral, unitary-matrix integration of 2D Yang-Mills action is presented. The uniqueness of the solution of heat equation enables us to integrate out the unitary-matrix parts of hermite matrices and to obtain the expression of integration over vectors, also in this case. 
\end{abstract}

\vspace{2.5cm}

\noindent {\bf Introduction}

\vskip 0.5cm

In 1980, C. Itzykson and J.-B. Zuber presented the method of integration over angular variables on a simplest model coupling two hermite matrices~\cite{itz}. Here, the angular variable means the unitary matrix which diagonalizes physical quantitiy, i.e. $N\! \times \! N$ hermite matrix. Their work had a great impact and enabled developments especially on the theories of random matrix and $2D$ gravity (~\cite{mehta} etc.). While, until then, there was Creutz's unitary-matrix integration~\cite{creutz} well known in lattice gauge theory which integrate unitary matrices order by order in high temperature expansion, Itzykson-Zuber integral is epoch-making in the sense that they obtained the expression only using the eigenvalues of hermite matrices, after exactly integrating out angular variables, to all orders of exponential. In practice, since it is too complicated to integrate in usual ways, they evaluate the integral with the partial differential equation known as heat equation whose uniqueness of its solution is completely understood. They showed that the expression after integration is equivalent to that before integration using the uniqueness of the solution. After the unitary-matrix integration, the number of variables is reduced to $N$, i.e. vector, from $N\! \times \! N$ matrix, and therefore, the analyses such as $N\! \to \! \infty$ planar limit becomes very easy.

In this paper, using their idea, we present a similar expression after integrating out unitary matrices on 2-dimensional Yang-Mills action with intent to prepare to apply this, in future, to 10-dimensional Yang-Mills theory, i.e. IIB Matrix Model~\cite{ikkt,ikkt2} which is known to be convergent~\cite{ymi}. On our model, if we introduce an external field as $N\! \times \! N$ hermite matrix, the same heat equation as Itzykson-Zuber integral is also found to be satisfied, and therefore, we can evaluate our integral on the basis of the uniqueness of its solution. However, unlike Itzykson-Zuber integral, since a number of coefficients which cannot be determined only from the uniqueness remain, we derive them from the structure of the commutator characteristic of Yang-Mills action. As a result, in case $N$ is even, after integrating out angular variables, we find that the partition function of 2-dimensional Yang-Mills action can be expressed in terms of the remaining eivenvalues.

\vskip 1.5cm

\noindent {\bf Unitary-Matrix Integration of 2D Yang-Mills Action}

\vskip 0.5cm

The integral we investigate is the Yang-Mills action in 2 ``space-time" dimensions 

\begin{eqnarray}
	I(C;g)=\int dAdB \> \exp \Big\{ g\> tr[A,B]^2+i\> trC[A,B] \Big\} ,
\end{eqnarray}

\noindent which exists in the ``external field" $C$. Here, $A,B$ and $C$ are $N\! \times \! N$ hermitian matrices, and $iC[A,B]$ is also hermitian. $dA$ (and $dB$) is the measure 

\begin{eqnarray}
	dA=\prod_{i=1}^N dA_{ii} \prod_{i<j}^N d(ReA_{ij})d(ImA_{ij}),
\end{eqnarray}

\noindent which is invariant under the adjoint operation of $U(N)$ matrices 

\begin{eqnarray*}
	d(V^{\dagger}AV)=dA, \qquad V\in U(N).
\end{eqnarray*}

\noindent This measure can be written in the diagonalized form as follows: 

\begin{eqnarray}
	dA=\prod_{i=1}^N da_i \> dU \> \Delta^2 (\vec{a}),
\end{eqnarray}

\noindent where $A$ is diagonalized by unitary matrix $U$ such as $A\! =\! U^{\dagger}aU, \> a\! =\! diag(a_1,\cdots ,a_N)$, and $dU$ is the normalized $U(N)$ Haar measure, $\int \! dU \! =\! 1$. $\Delta (\vec{a})$ is the Vandermonde determinant 

\begin{eqnarray}
	\Delta (\vec{a})=(-1)^{\frac{N(N-1)}{2}}
	\left| \begin{array}{cccc}
	1 & 1 & \cdots & 1 \\
	a_1 & a_2 & \cdots & a_N \\
	a_1^2 & a_2^2 & \cdots & a_N^2 \\
	\vdots & \vdots & \ddots & \vdots \\
	a_1^{N-1} & a_2^{N-1} & \cdots & a_N^{N-1}
	\end{array} \right|
	=\prod_{i<j}^N (a_i-a_j).
\end{eqnarray}

As we can check easily, the Itzykson-Zuber integral~\cite{itz}, 

\begin{eqnarray}
	K(C;g) \! \! \! & = & \! \! \! \int dX \exp \Big\{ -g\> trX^2 
	+\> trCX \Big\} \nonumber \\
	& = & \! \! \! \int d^N\vec{x} \> 
	\frac{\Delta (\vec{x})}{\Delta (\vec{c})} \> 
	\exp \Big\{ -g|\vec{x}|^2+\vec{c}\cdot \vec{x} \Big\},
\end{eqnarray}

\noindent obeys the following partial differenial equation, 

\begin{eqnarray}
	& & tr\Big( \frac{\partial^2}{\partial C^2}\Big) K(C;g)
	=-\frac{\partial}{\partial g} K(C;g), \\
	& & \quad tr\Big( \frac{\partial^2}{\partial C^2}\Big) \equiv 
	\sum_{i,j=1}^N \frac{\partial}{\partial C_{ij}}
	\frac{\partial}{\partial C_{ji}}. \nonumber
\end{eqnarray}

\noindent Essentially, this equation is the heat equation, and has an unique solution (5) under the boundary condition $K(C;g\! \to \! \infty)\! =\! 0$.

Similarly, the integral (1) satisfies the same equation 

\begin{eqnarray}
	tr\Big( \frac{\partial^2}{\partial C^2}\Big) I(C;g)
	=-\frac{\partial}{\partial g} I(C;g),
\end{eqnarray}

\noindent and must have an unique solution under the boundary condition $I(C;g\! \to \! \infty)\! =\! 0$. After integrating over $U(N)$ matrices, we expect that (1) becomes the integral which only contains the integration over two eigenvalues of hermite matrices $A$ and $B$. Then, we suppose the integral (1) to become like the following form: 

\begin{eqnarray}
	i(\vec{c};g)=\int d^N\vec{a} d^N\vec{b} \> \exp \Big\{ 
	-g\sum_{i,j,k,l=1}^N R_{ijkl} a_ib_ja_kb_l +\sum_{i,j,k=1}^N 
	\Gamma_{ijk} c_ia_jb_k \Big\}.
\end{eqnarray}

\noindent Here, the coefficients $R_{ijkl}$ and $\Gamma_{ijk}$ are independent of $\vec{c}\! =\! (c_1,\cdots ,c_N)$. Since this integral must obey the partial differential equation 

\begin{eqnarray}
	\sum_{j=1}^N \frac{\partial^2}{\partial c_j^2} i(\vec{c};g)
	=-\frac{\partial}{\partial g} i(\vec{c};g),
\end{eqnarray}

\noindent $R_{ijkl}$ is determined from $\Gamma_{ijk}$: 

\begin{eqnarray}
	R_{ijkl}=\sum_{m=1}^N \Gamma_{mij}\Gamma_{mkl}.
\end{eqnarray}

\noindent Therefore, it is sufficient we obtain $\Gamma_{ijk}$, and, from the equation 

\begin{eqnarray}
	& & tr\Big( \frac{\partial^2}{\partial C^2}\Big)=\Delta^{-1}(\vec{c})
	\sum_{j=1}^N\frac{\partial^2}{\partial c_j^2}\> \Delta (\vec{c})
	+\nabla_{U_C}^2, \\
	& & \quad C=U_C^{\dagger}cU_C, \nonumber
\end{eqnarray}

\noindent where $\nabla_{U_C}^2$ only depends on $U_C$, the integral (1) becomes 

\begin{eqnarray}
	I(C;g)=\int d^N\vec{a} d^N\vec{b} \> 
	\frac{f(\vec{a},\vec{b})}{\Delta(\vec{c})} \> \exp \Big\{ 
	-g\sum_{i,j,k,l=1}^N R_{ijkl} a_ib_ja_kb_l +\sum_{i,j,k=1}^N 
	\Gamma_{ijk} c_ia_jb_k \Big\},
\end{eqnarray}

\noindent owing to the uniqueness of the solution of (7). Here, $f(\vec{a},\vec{b})$ is the undetermined function of $\vec{a}$ and $\vec{b}$. Furthermore, though $\vec{c}$ is exactly the eigenvalue of hermite matrix $C$, we cannot certain that both $\vec{a}$ and $\vec{b}$ are those of $A$ and $B$, yet. However, with $g\! =\! 0$ in (12), the integral 

\begin{eqnarray}
	\int dUdV \! \! \! \! \! \! \! \! & & \exp \Big\{ i\> 
	trC[U^{\dagger}AU,V^{\dagger}BV] \Big\} \nonumber \\
	& & = \int dUdVdW \> \exp \Big\{ i\> trW^{\dagger}CW[U^{\dagger}AU,
	V^{\dagger}BV] \Big\} \nonumber \\
	& & = \int dVdW \> \exp \Big\{ i\> trA[V^{\dagger}BV,
	W^{\dagger}CW] \Big\} \nonumber \\
	& & = \int dUdW \> \exp \Big\{ i\> trB[W^{\dagger}CW,
	U^{\dagger}AU] \Big\} \nonumber \\
	& & = \frac{1}{\Delta (\vec{a})\Delta (\vec{b})\Delta (\vec{c})}\exp 
	\Big\{ \sum_{i,j,k=1}^N \Gamma_{ijk} c_ia_jb_k \Big\}
\end{eqnarray}

\noindent enables us to derive $f(\vec{a},\vec{b})\! =\! \Delta (\vec{a})\Delta (\vec{b})$ and make sure that both $\vec{a}$ and $\vec{b}$ are the eigenvalues of $A$ and $B$, because all $A,B$ and $C$ stand on the same ground. And, also from the facts that $trC[A,B]=trA[B,C]=trB[C,A]$ and $trC[A,B]\! =\! -trC[B,A]$, the index of $\Gamma_{ijk}$ has cyclic symmetry: $\Gamma_{ijk}\! =\! \Gamma_{jki}\! =\! \Gamma_{kij}$ and anti-symmetry under the interchange of its indices: $\Gamma_{ijk}\! =\! -\Gamma_{ikj}$. Therefore, we obtain 

\begin{eqnarray}
	I(C;g) \! \! \! \! \! \! \! \! 
	& & =\int dAdB \> \exp \Big\{ g\> tr[A,B]^2+i\> trC[A,B] \Big\} 
	\nonumber \\
	& & =\int d^N\vec{a} d^N\vec{b} \> 
	\frac{\Delta (\vec{a})\Delta (\vec{b})}{\Delta(\vec{c})} \> \exp 
	\Big\{ -g\sum_{i,j,k,l=1}^N R_{ijkl} a_ib_ja_kb_l +\sum_{i,j,k=1}^N 
	\Gamma_{ijk} c_ia_jb_k \Big\}. \quad
\end{eqnarray}

The remaining task we must do is to determine the coefficients $\Gamma_{ijk}$. Since to have the matrices $A$ and $B$ multiplied by the constants $k$ is equivalent to make the eigenvalues $k\vec{a}$ and $k\vec{b}$ respectively, $\Gamma_{ijk}$ are also independent of $\vec{a}$ and $\vec{b}$. Then, we can evaluate $\Gamma_{ijk}$ through the structure peculiar to the commutator $[\ast ,\ast ]$. We can easily check the integral 

\begin{eqnarray}
	I(\vec{a},\vec{b};g) \! \! \! \! \! \! \! \! 
	& & =\int dU_AdU_B \> \exp \Big\{ 
	g\> tr[U_A^{\dagger}aU_A,U_B^{\dagger}bU_B]^2 \Big\} \\
	& & =\exp \Big\{ -g\sum_{i,j,k,l=1}^N R_{ijkl} a_ib_ja_kb_l \Big\}
\end{eqnarray}

\noindent follows from (14), and (15) satisfies the differential equation due to the characteristic structure of the commutator: 

\begin{eqnarray}
	\sum_{i=1}^N \frac{\partial}{\partial a_i}I(\vec{a},\vec{b};g)=0, 
\end{eqnarray}

\noindent therefore, from (16), we obtain the algebraic equation which $R_{ijkl}$ must obey: 

\begin{eqnarray}
	\sum_{i,j,k,l=1}^N R_{ijkl}b_ja_kb_l
	=\sum_{i,j,k,l,m=1}^N \Gamma_{mij}\Gamma_{mkl}b_ja_kb_l=0.
\end{eqnarray}

\noindent This is the identity about $a_i$ and $b_i$. By introducing $d_m\! \equiv \! \sum_{k,l}\Gamma_{mkl}a_kb_l$, this reduces to the identity 

\begin{eqnarray}
	\sum_{i,j,m=1}^N \Gamma_{mij}b_jd_m=0
\end{eqnarray}

\noindent about $b_j$ and $d_m$, and we obtain conditions for $\Gamma_{ijk}$: 

\begin{eqnarray}
	\sum_{i=1}^N \Gamma_{ijk}=0, \qquad \text{for fixed }j,k.
\end{eqnarray}

\noindent When $N\! =\! 4$, these equations become 

\begin{eqnarray*}
	& & \Gamma_{312}+\Gamma_{412}=0,\quad \Gamma_{123}+\Gamma_{423}=0, \\
	& & \Gamma_{213}+\Gamma_{413}=0,\quad \Gamma_{124}+\Gamma_{324}=0, \\
	& & \Gamma_{214}+\Gamma_{314}=0,\quad \Gamma_{134}+\Gamma_{234}=0,
\end{eqnarray*}

\noindent then, except overall factor, all $\Gamma_{ijk}$s are determined exactly as 

\begin{eqnarray}
	\Gamma_{123}=-\Gamma_{124}=\Gamma_{134}=-\Gamma_{234}=1.
\end{eqnarray}

\noindent When $N\! =\! 2$ and $N\! =\! 3$, $\Gamma_{ijk}\! =\! 0$ for all $i,j,k$. When $N\! \geq \! 4$, from the symmetry of its indices, $\Gamma_{ijk}$ has ${}_NC_3$ independent components, but (20) are ${}_NC_2$ equations at most. So, we cannot determine all $\Gamma_{ijk}$ from (20) in general. However, as is obvious from (14), all $a_i,b_i$ and $c_i$ are contained in the same weight. That is, except the overall factor which can be absorbed into the integration measure, we can set $\big|\Gamma_{ijk}\big|\! =\! 1$. Then, for even $N\! \geq \! 4$, we can find the solution of (20): 

\begin{eqnarray}
	\Gamma_{ijk}=(-1)^{i+j+k}, \qquad \text{for} \quad i<j<k.
\end{eqnarray}

\noindent When $N$ is odd, the solution such as $\big|\Gamma_{ijk}\big|\! =\! 1$ does not exist, because odd numbers of $\pm 1$ cannot add up to zero. In fact, when $N\! =\! 5$, 

\begin{eqnarray*}
	& & \Gamma_{312}+\Gamma_{412}+\Gamma_{512}=0,\quad 
	\Gamma_{124}+\Gamma_{324}+\Gamma_{524}=0, \\
	& & \Gamma_{213}+\Gamma_{413}+\Gamma_{513}=0,\quad 
	\Gamma_{125}+\Gamma_{325}+\Gamma_{425}=0, \\
	& & \Gamma_{214}+\Gamma_{314}+\Gamma_{514}=0,\quad 
	\Gamma_{134}+\Gamma_{234}+\Gamma_{534}=0, \\
	& & \Gamma_{215}+\Gamma_{315}+\Gamma_{415}=0,\quad 
	\Gamma_{135}+\Gamma_{235}+\Gamma_{435}=0, \\
	& & \Gamma_{123}+\Gamma_{423}+\Gamma_{523}=0,\quad 
	\Gamma_{145}+\Gamma_{245}+\Gamma_{345}=0,
\end{eqnarray*}

\noindent the solution does not exist. This strange feature that we can use the expression (14) only for even $N$ seems to come from the following characteristic of the commutator $[\ast ,\ast ]$. The algebra of hermite matrices has the subalgebra which consists of real symmetric matrices $A^{\dagger}\! =\! A^T\! =\! A$, and, in this subalgebra, the commutator $[A,B]$ always becomes the anti-symmetric matrix $[A,B]^{T}\! =\! -[A,B]$ which has even rank. That is, the commutator $[\ast ,\ast ]$ essentially contains the structure of even rank which is hidden by larger $U(N)$ symmetry before the integration, and then, the integration of $U(N)$ matrices may reveal this structure. However, since we need the integral after taking the limit $N\! \to \! \infty$, it is sufficient for our purpose to have the result that $N$ is even number.

Lastly, let us line up the result we obtain in this paper.

\begin{eqnarray*}
	I(C;g) \! \! \! \! \! \! \! \! 
	& & =\int dAdB \> \exp \Big\{ g\> tr[A,B]^2+i\> trC[A,B] \Big\} 
	\nonumber \\
	& & =\int d^N\vec{a} d^N\vec{b} \> 
	\frac{\Delta (\vec{a})\Delta (\vec{b})}{\Delta(\vec{c})} \> \exp 
	\Big\{ -g\sum_{i,j,k,l=1}^N R_{ijkl} a_ib_ja_kb_l +\sum_{i,j,k=1}^N 
	\Gamma_{ijk} c_ia_jb_k \Big\}, \\
	R_{ijkl} \! \! \! \! \! \! \! \! 
	& & =\sum_{m=1}^N \Gamma_{mij}\Gamma_{mkl}, \\
	\Gamma_{ijk} \! \! \! \! \! \! \! \! 
	& & =\Gamma_{kij}=-\Gamma_{ikj}, \\
	\Gamma_{ijk} \! \! \! \! \! \! \! \! 
	& & =(-1)^{i+j+k}, \qquad \text{for} \quad i<j<k.
\end{eqnarray*}

\vskip 1.5cm

\noindent
{\bf Discussions}

\vskip 0.5cm

Using the uniqueness of the solution for heat equation which plays a crucial role on Itzykson-Zuber integral, we derive an expression which contains eigenvalues of hermite matrices only as integration variables after integrationg out angular variables on 2-dimensional Yang-Mills action. Then, characteristic structures of the commutator are important and needed to determine coefficients in the expression. And the expression is possible only for even $N$, reflecting the structures.

Our expression can be applied to various problems. For example, as the model of Itzykson-Zuber integral which can be solved exactly in $N\! \to \! \infty$ planar limit, we can evaluate similarly on 2-dimensional Yang-Mills model. And, as refered in introduction, by extending our result to larger dimensions, for example IIB Matrix Model, we may be able to treat the phenomenon of the phase-transition of space-time, i.e. inflation. However, when we extend to higher dimensions than three, we cannot probably use the uniqueness of the solution of heat equation, so must introduce other methods.

Furthermore, since Vandermonde determinant $\Delta (\vec{x})$, which always emerges in case of integrating over hermite matrices, is the wave function of ground state of Calogero-Sutherland model describing the system of fermions on circle~\cite{cs}, and the orthogonal polynomials containing $\Delta (\vec{x})$ satisfy particular recursion relations~\cite{itz}, we may be able to calculate exactly over remaining eigenvalues.

\vskip 1cm

\noindent \underline{ Acknowledgements }

\vskip 0.2cm

I would like to thank my colleagues for useful comments and discussions. Especially, I gratefully acknowledge S. Matsuura and S. Shinohara for helpful suggestions.


\end{document}